\documentclass[prl,12pt,showpacs]{revtex4}

\usepackage{dcolumn}
\usepackage{amssymb}
\usepackage{amsmath}
\usepackage{graphicx}

\begin{document}
\title{Charge and Spin Effects in Mesoscopic Josephson Junctions}

\author{Ilya V. Krive$^{1,2}$, Sergei I. Kulinich$^{1,2}$, Robert I. Shekhter$^1$ and
Mats Jonson$^1$ }
\affiliation{ $^{(1)}$Department of Applied
Physics, Chalmers University of Technology and G\"{o}teborg
University, SE-412 96 G\"{o}teborg, Sweden }
\affiliation{$^{(2)}$B.I. Verkin Institute for Low Temperature
Physics and Engineering, 47 Lenin Avenue, 61103 Kharkov, Ukraine}

\begin{abstract}
We consider the charge and spin effects in low dimensional
superconducting weak links. The first part of the review deals
with the effects of electron-electron interaction in
Superconductor/Luttinger liquid/Superconductor junctions. The
experimental realization of this mesoscopic hybrid system can be
the individual single wall carbon nanotube that bridges the gap
between two bulk superconductors. The dc Josephson current through
a Luttinger liquid in the limits of perfectly and poorly
transmitting junctions is evaluated. The relationship between the
Josephson effect in a long SNS junction and the Casimir effect is
discussed. In the second part of the paper we review the recent
results concerning the influence of the Zeeman and Rashba
interactions on the thermodynamical properties of ballistic S/QW/S
junction fabricated in two dimensional electron gas. It is shown
that in magnetically controlled junction there are conditions for
resonant Cooper pair transition which results in giant
supercurrent through a tunnel junction and a giant magnetic
response of a multichannel SNS junction. The supercurrent induced
by the joint action of the Zeeman and Rashba interactions in 1D
quantum wires connected to bulk superconductors is predicted.
\end{abstract}


\maketitle
\newpage
\begin{center}
{\bf Contents}
\end{center}
 1. Introduction \\
 2. Josephson current in  S/QW/S junction \\
 \hspace*{.2in}  2.1. Quantization of Josephson current in a short ballistic junction \\
 \hspace*{.2in} 2.2. Luttinger liquid wire coupled to superconductors \\
 \hspace*{.5in}  2.2.1. Tunnel junction\\
 \hspace*{.5in} 2.2.2. Transparent junction \\
 \hspace*{.2in} 2.3. Josephson current and the Casimir effect \\
3. The effects of Zeeman splitting and spin-orbit interaction in
   SNS junctions \\
\hspace*{.2in}3.1. Giant critical current in a magnetically controlled tunnel junction \\
\hspace*{.2in}3.2. Giant magnetic response of a multichannel quantum wire coupled
to superconductors \\
\hspace*{.2in}3.3. Rashba effect and chiral electrons in quantum wires \\
\hspace*{.2in}3.4. Zeeman splitting induced supercurrent \\
4. Conclusion \\
\hspace*{.2in}References

\section{1. Introduction}
Since the discovery of superconductivity in 1911 this amazing
macroscopic quantum phenomena has influenced modern solid state
physics more then any other fundamental discovery in the 20th
century. The mere fact that five Nobel prizes already have been
awarded for discoveries directly connected to superconductivity
indicates the worldwide recognition of the exceptional role
superconductivity plays in physics.

Both at the early stages of the field development and later on,
research in basic superconductivity brought surprises. One of the
most fundamental discoveries made in superconductivity was the
Josephson effect \cite{i6}. In 1962 Josephson predicted that when
two superconductors are put into contact via an insulating layer
(SIS junction) then (i) a dc supercurrent $J=J_c\sin \varphi$
($J_c$ is the critical current, $\varphi$ is the superconducting
phase difference) flows through the junction in equilibrium (dc
Josephson effect) and (ii) an alternating current
($\varphi=\omega_J t, \, \omega_J=2eV/\hbar,$ where $V$ is the
bias voltage) appears when a voltage is applied across the
junction (ac Josephson effect). A year latter both the dc and the
ac Josephson effect were observed in experiments \cite{i7, i8}. An
important contribution to the experimental proof of the Josephson
effect has been made by Yanson, Svistunov and Dmitrenko \cite{i9},
who were the first to observe rf-radiation from the voltage biased
contact and who measured the temperature dependence of the
critical Josephson current $J_c(T)$.

As a matter of fact the discovery of the Josephson effect gave
birth to a new and unexpected direction in superconductivity,
namely, the superconductivity of weak links (weak
superconductivity, see e.g.Ref.~[\onlinecite{i10}]). It soon
became clear that any normal metal layer between superconductors
(say, an SNS junction) will support a supercurrent as long as the
phase coherence in the normal part of the device is preserved.
Using the modern physical language one can say that the physics of
superconducting weak links turned out to be part of mesoscopic
physics.

During the last decade the field of mesoscopic physics has been
the subject of an extraordinary growth and development. This was
mainly caused by the recent advances in fabrication technology and
by the discovery of principally new types of mesoscopic systems
such as carbon nanotubes (see e.g.Ref.~[\onlinecite{i11}]).

For our purposes metallic single wall carbon nanotubes (SWNT) are
of primary interest since they are strictly one-dimensional
conductors. It was experimentally demonstrated \cite{i12, i13,
i14} (see also Ref.~[\onlinecite{i15}]) that electron transport
along metallic individual SWNT at the low bias voltage regime is
ballistic. At first glance this observation looks surprising. For
a long time it was known (see e.g. Ref.~[\onlinecite{i15a}]) that
1D metals are unstable with respect to the Peierls phase
transition, which opens up a gap in the electron spectrum at the
Fermi level. In carbon nanotubes the electron-phonon coupling for
conducting electrons is very weak while the Coulomb correlations
are strong. The theory of metallic carbon nanotubes \cite{i16,
i17} shows that at temperatures outside the $m K$ range the
individual SWNT has to demonstrate the properties of a two
channel, spin-1/2 Luttinger liquid (LL). This theoretical
prediction was soon confirmed by transport measurements on
metal-SWNT and SWNT-SWNT junctions \cite{i17,i18,i19} (see also
Ref.~[\onlinecite{i20}], where the photoemission measurements on a
SWNT were interpreted as a direct observation of LL state in
carbon nanotubes). Both theory and experiments revealed strong
electron-electron correlations in SWNTs.

Undoped individual SWNT is not intrinsically a superconducting
material. Intrinsic superconductivity was observed only in ropes
of SWNT (see Refs.~[\onlinecite{i21, i22}]). Here we consider the
proximity-induced superconductivity in a LL wire coupled to
superconductors (SLLS). The experimental realization of SLLS
junction could be an individual SWNT, which bridges the gap
between two bulk superconductors \cite{i23, i24}.

The dc Josephson current through a LL junction was evaluated for
the first time in Ref.~[\onlinecite{220}]. In this paper a tunnel
junction was considered in the geometry (see subsection 2.2.),
which is very suitable for theoretical calculations but probably
difficult to realize in an experiment. It was shown that the
Coulomb correlations in a LL wire strongly suppress the critical
Josephson current. The opposite limit - a perfectly transmitting
SLLS junction was studied in Ref.~[\onlinecite{225}], where it was
demonstrated by a direct calculation of the dc Josephson current
that the interaction does not renormalize the supercurrent in a
fully transparent ($D=1$, $D$ is the junction transparency)
junction. In subsection 2.2. we re-derive and explain these
results using the boundary Hamiltonian method \cite{213}.

The physics of quantum wires is not reduced to the investigations
of SWNTs. Quantum wires can be fabricated in a two-dimensional
electron gas (2DEG) by using various experimental methods. Some of
them (e.g. the split-gate technique) originate from the end of
80's when the first transport experiments with a quantum point
contact (QPC) revealed unexpected properties of quantized electron
ballistic transport (see e.g. Ref.~[\onlinecite{i25}]). In
subsection 2.1. we briefly review the results concerning the
quantization of the critical supercurrent in a QPC.

In quantum wires formed in a 2DEG the electron-electron
interaction is less pronounced \cite{i26} than in SWNTs
(presumably due to the screening effects of nearby bulk metallic
electrodes). The electron transport in these systems can in many
cases be successfully described by Fermi liquid theory. For
noninteracting quasiparticles the supercurrent in a SNS ballistic
junction is carried by Andreev levels. For a long ($L\gg
\xi_0=\hbar v_F/\Delta,\, L$ is the junction length, $\Delta$ is
the superconducting energy gap) perfectly transmitting junction
the Andreev-Kulik spectrum \cite{26} for quasiparticle energies
$E\ll \Delta$ is a set of equidistant levels. In subsection 2.3.
we show that this spectrum corresponds to twisted periodic
boundary conditions for chiral (right- and left-moving) electron
fields and calculate the thermodynamic potential of an SNS
junction using field theoretical methods. In this approach there
is a close connection between the Josephson effect and the Casimir
effect.

In section 3 of our review we consider the spin effects in
ballistic Josephson junctions. As is well-known, the electron spin
does not influence the physics of standard SIS or SNS junctions.
Spin effects become significant for SFS junctions (here ''F"
denotes a magnetic material) or when spin-dependent scattering on
magnetic impurities is considered. As a rule, magnetic impurities
tend to suppress the critical current in Josephson junction by
inducing spin-flip processes \cite{i27, i28}. Another system where
spin effects play an important role is a quantum dot (QD).
Intriguing new physics appears in normal and superconducting
charge transport through a QD at very low temperatures when the
Kondo physics starts to play a crucial role in the electron
dynamics. Last year a vast literature was devoted to these
problems.

Here we discuss the spin effects in a ballistic SNS junction in
the presence of: (i) the Zeeman splitting due to a local magnetic
field acting only on the normal part of the junction, and (ii)
strong spin-orbit interaction, which is known to exist in quantum
heterostructures due to the asymmetry of the electrical confining
potential \cite{315}. It is shown in subsection 3.1. that in
magnetically controlled single barrier junction there are
conditions when superconductivity in the leads strongly enhances
electron transport, so that a giant critical Josephson current
appears $J_c\sim \sqrt{D},$ ($D$ is the junction transparency).
The effect is due to resonant electron transport through de
Gennes-Saint-James energy levels split by tunneling.

The joint action of Zeeman splitting and superconductivity (see
subsection 3.2.) results in yet another unexpected effect - a
giant magnetic response, $M\sim N_\bot \mu_B$, ($M$ is the
magnetization, $N_\bot$ is the number of transverse channels of
the wire, $\mu_B$ is the Bohr magneton) of a multichannel quantum
wire coupled to superconductors \cite{235}. This effect can be
understood in terms of the Andreev level structure which gives
rise to an additional (superconductivity-induced) contribution to
the magnetization of the junction. The magnetization peaks at
special values of the superconducting phase difference when the
Andreev energy levels at $E_\pm = \pm \Delta_Z$, ($\Delta_Z$ is
the Zeeman energy splitting) become $2N_\bot$-fold degenerate.

The last two subsections of section 3 deal with the influence of
the Rashba effect on the transport properties of quasi-1D quantum
wires. Strong spin-orbit (s-o) interaction experienced by 2D
electrons in heterostructures in the presence of additional
lateral confinement results in a dispersion asymmetry of the
electron spectrum in a quantum wire and in a strong correlation
between the direction of electron motion along the wire
(right/left) and the electron spin projection \cite{316, 323}.

The chiral properties of electrons in a quantum wire cause
nontrivial effects when the wire is coupled to bulk
superconductors. In particular, in subsection 3.4. we show that
the Zeeman splitting in a S/QW/S junction induces an anomalous
supercurrent, that is a Josephson current that persists even at
zero phase difference between the superconducting banks.

In the Conclusion we once more emphasize the new features of the
Josephson current in ballistic mesoscopic structures and briefly
discuss the novel effects, which could appear in an ac Josephson
current through an ultra-small superconducting quantum dot.

\section{2. Josephson current through a superconductor/
quantum wire/superconductor junction}

In this chapter we consider the Josephson current in a quantum
wire coupled to bulk superconductors. One could expect that the
conducting properties of this system strongly depend on the
quality of the electrical contacts between the QW and the
superconductors. The normal conductance of a QW coupled to
electron reservoirs in Fermi liquid theory is determined by the
transmission properties of the wire (see e.g.
Ref.[\onlinecite{21}]). For the ballistic case the transmission
coefficient of the system in the general situation of nonresonant
electron transport depends only on the transparencies of the
potential barriers which characterize the electrical contacts and
does not depend on the length $L$ of the wire. As already was
mentioned in the Introduction, the Coulomb interaction in a long
1D (or few transverse channel) QW is strong enough to convert the
conduction electrons in the wire into a Luttinger liquid. Then the
barriers at the interfaces between QW and electron reservoirs are
strongly renormalized by electron-electron interaction and the
conductance of the N/QW/N junction at low temperature strongly
depends on the wire length \cite{22}. For a long junction and
repulsive electron-electron interaction the current through the
system is strongly suppressed. The only exception is the case of
perfect (adiabatic) contacts when the backscattering of electrons
at the interfaces is negligibly (exponentially) small. In the
absence of electron backscattering the conductance $G$ is not
renormalized by interaction \cite{23} and coincides with the
conductance quantum $G=2e^2/h$ (per channel). From the theory of
Luttinger liquis it is also known \cite{24} that for a strong
repulsive interaction the resonant transition of electrons through
a double-barrier structure is absent even for symmetric barriers.

The well-known results for the transport properties of 1D
Luttinger liquid listed above (see e.g. review paper \cite{25})
allows us to consider two cases when studing ballistic S/QW/S
junctions: (i) a transparent junction ($D=1$, $D$ is the junction
transparency) and (ii) a tunnel junction ($D\ll 1$). These two
limiting cases are sufficient to describe the most significant
physical effects in S/QW/S junctions.

\subsection{2.1. Quantization of the Josephson Current
in a Short Ballistic Junction}

At first we consider a short $L\ll \xi_0\; (\xi_0=\hbar
v_F/\Delta$ is the coherence length and $\Delta$ is the
superconducting gap) ballistic  S/QW/S junction. One of the
realizations of this  mesoscopic device is a quantum point contact
(QPC) in a 2DEG (see Fig.~1a). For a QPC the screening of the
Coulomb interaction is qualitatively the same as in a pure 2D
geometry and one can evaluate the Josephson current through the
constriction in a noninteracting electron model. Then due to
Andreev backscattering of quasiparticles at the SN interfaces, a
set of Andreev levels is formed in the normal part of the junction
\cite{26}. In a single mode short junction the spectrum of bound
states takes the form \cite{27} ($L/\xi_0\rightarrow 0$)
\begin{equation} \label{21}
E_\pm=\pm \Delta \sqrt{1-D\sin^2 \varphi/2}\,,
\end{equation}
where $\varphi$ is the superconducting phase difference. This
spectrum does not depend on the Fermi velocity and therefore the
Andreev levels, Eq.~(\ref{21}),in a junction with $N_\bot$
transverse channels are $2N_\bot$ degenerate (the factor $2$ is
due to spin degeneracy).

It is well known (see e.g. Ref.~[\onlinecite{29, 226}]) that
the continuum spectrum in the limit
$L/\xi_0\rightarrow 0$ does not contribute to the Josephson
current,
\begin{equation}\label{22}
J=\frac{e}{\hbar} \frac{\partial \Omega}{\partial \varphi}\,,
\end{equation}
where $\Omega$ is the thermodynamic potential. It is evident from
Eqs. (\ref{21}) and (\ref{22}) that the Josephson current through
a QPC ($D=1$) is quantized \cite{29}. At low temperatures ($T\ll
\Delta$) we have \cite{29}
\begin{equation}\label{23}
J=N_\bot \frac{e \Delta}{\hbar}\sin\frac{\varphi}{2} \,.
\end{equation}
This effect (still not observed experimentally) is the analog of
the famous conductance quantization in OPCs (see
Ref.~[\onlinecite{210}]).

Now let us imagine that the geometry of the constriction allows
one to treat the QPC as a 1D quantum wire of finite length $L$
smoothly connected to bulk superconductors (Fig.~1b). The 1D wire
is still much shorter that the coherence length $\xi_0$. How does
the weakly screened Coulomb interaction in a 1D QW influence the
Josephson current in a fully transmitting ($D=1$) junction? Notice
that the charge is freely transported through the junction since
the real electrons are not backscattered by the adiabatic
constriction \cite{99}. So, it is reasonable to assume that the
Coulomb interaction in this case does not influence the Josephson
current at all. We will prove this assumption for the case of a
long junction in the next section.
\begin{figure}
\begin{center}
\includegraphics[angle=0, width=15cm]{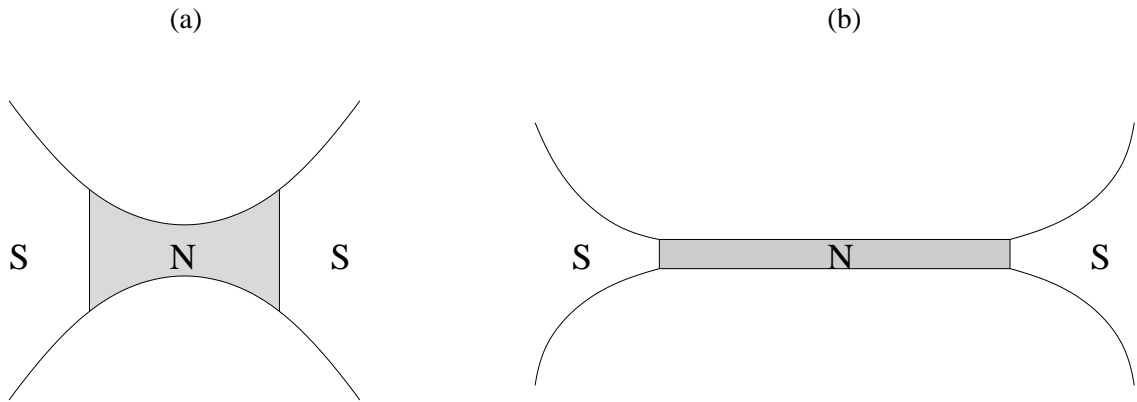}
\caption{(a) A schematic display of a superconducting point
contact. (b) Quantum wire adiabatically connected to bulk
superconductors.}
\end{center}
\end{figure}
If the QW is separated from the leads by potential barriers (quite
a natural situation in a real experiment) the charging effects
have to be taken into account. As a rule the Coulomb correlations,
which tend to keep the number of electrons in the normal region
(quantum dot in our case) constant, suppress the critical
supercurrent due to the Coulomb blockade effect (see e.g.
Ref.~[\onlinecite{211}], where a consistent theory of the Coulomb
blockade of Josephson tunneling was developed). They can also
change the $\varphi$-dependence of the Josephson current. One
possible scenario for how charging effects influence the Josephson
current in a short SNS junction is considered in
Ref.~[\onlinecite{212}].

\subsection{2.2. Luttinger Liquid Wire Coupled to Superconductors}

A consistent theory of electron-electron interactions effects in
weak superconductivity has been developed for a long 1D or
quasi-1D SNS junction, when the normal region can be modelled by a
Luttinger liquid (LL). The standard approach to this problem (see
e.g. Ref.~[\onlinecite{213}]) is to use for the description of
electron transport through the normal region the LL Hamiltonian
with boundary conditions which take into account the Andreev
\cite{227} and normal backscattering of quasiparticles at the NS
interfaces.

The LL Hamiltonian $H_{LL}$ expressed in terms of charge density
operators $ \tilde{\rho}_{R/L,\uparrow / \downarrow}$ of
right/left moving electrons with up/down spin projection takes the
form (see e.g. Ref.~[\onlinecite{214}])
\begin{eqnarray}\label{24}
H_{LL} &=& \pi\hbar\int dx [u(\tilde{\rho}_{R\uparrow}^2 +
\tilde{\rho}_{L\uparrow}^2 + \tilde{\rho}_{R\downarrow}^2
+\tilde{\rho}_{L\downarrow}^2) + \nonumber \\
&+& \frac{V_0}{\pi\hbar}(\tilde{\rho}_{R\uparrow}
\tilde{\rho}_{R\downarrow} + \tilde{\rho}_{L\uparrow}
\tilde{\rho}_{L\downarrow} +
\tilde{\rho}_{R\uparrow}\tilde{\rho}_{L\uparrow} +
\tilde{\rho}_{R\downarrow} \tilde{\rho}_{L\downarrow} +
\tilde{\rho}_{R\uparrow}\tilde{\rho}_{L\downarrow} +
\tilde{\rho}_{R\downarrow} \tilde{\rho}_{L\uparrow})]\,,
\end{eqnarray}
where $V_0$ is the strength of electron-electron interaction
($V_0\sim e^2$) and the velocity $u=v_F+V_0/2\pi\hbar$. The charge
density operators of the chiral (R/L) fields obey anomalous
Kac-Moody commutation relations (see e.g. Ref.~[\onlinecite{214}])
$$[\tilde{\rho}_{R(L)j}(x), \tilde{\rho}_{R(L)k}(x')]=\pm
\frac{\delta_{jk}}{2\pi\imath}\frac{\partial}{\partial x}
\delta(x-x')\;\;\;,\;\;\; j,k=\uparrow,\downarrow$$
The Hamiltonian (\ref{24}) is quadratic and can easily be
diagonalized by a Bogoliubov transformation
\begin{equation}\label{25}
H_{LL}^{(d)}=\pi\hbar\int dx \left[v_\varrho \left(
\rho^2_{R\varrho}+\rho^2_{L\varrho}\right) + v_\sigma \left(
\rho^2_{R\sigma}+\rho^2_{L\sigma}\right)\right]\,,
\end{equation}
where $v_{\varrho (\sigma)}$ are the velocities of noninteracting
bosonic modes (plasmons), $v_{\varrho(\sigma)}=v_F/g_{\varrho
(\sigma)}$, and
\begin{equation}\label{26}
g_\varrho =\left( 1+\frac{2V_0}{\pi\hbar v_F}\right)^{-1/2}\,,
\:g_\sigma =1\,.
\end{equation}
Here $g_\varrho$ and $g_\sigma$ are the correlation parameters of
a spin-1/2 LL in the charge $(\varrho)$ and spin $(\sigma)$
sectors. Notice that $g_\varrho\ll 1$ for a strongly interacting
($V_0\gg \hbar v_F$) electron system.

The Andreev and normal backscattering of quasiparticles at the NS
boundaries ($x=0$ and $x=L$) can be represented by the effective
boundary Hamiltonian $H_B=H_B^{(A)}+H_B^{(N)}$
\begin{eqnarray}
H_B^{(A)}&=&\Delta_B^{(l)}\left[ \Psi_{R \uparrow} (0)
\Psi_{L\downarrow} (0)+ \Psi_{R \downarrow} (0)
\Psi_{L\uparrow}(0) \right]+ \label{27} \\
&+&\Delta_B^{(r)} \left[ \Psi_{R \uparrow} (L) \Psi_{L\downarrow}
(L)+ \Psi_{R \downarrow}(L)
\Psi_{L\uparrow}(L) \right] + h.c.\,, \nonumber \\
H_B^{(N)}&=& V_B^{(l)}\sum_{j, \sigma} \Psi_{j\sigma}^\dag (0)
\Psi_{j\sigma}(0)+ V_B^{(r)} \sum_{j,\sigma} \Psi_{j\sigma}^\dag
(L) \Psi_{j\sigma}(L) \label{28}\,,
\end{eqnarray}
where $j=(L,R), \sigma=(\uparrow, \downarrow)$. Here
$\Delta_B^{(l,r)}$ is the effective boundary pairing potential at
the left (right) NS interface and $V_B^{(l,r)}$ is the effective
boundary scattering potential. The values of these potentials are
related to the phase of the superconducting order parameters in
the banks and to the normal scattering properties at the left and
right interfaces. They can be considered either as input
parameters (see e.g. Ref.~[\onlinecite{215}]) or they can be
calculated by using some particular model of the interfaces
\cite{213}. In what follows we will consider two limiting cases:
(i) poorly transmitting interfaces $V_B^{(l,r)}\rightarrow \infty$
(tunnel junction) and (ii) perfectly transmitting interfaces
$V_B^{(l,r)}\rightarrow 0$.

At first we relate the effective boundary pairing potentials
$\Delta_B^{(l,r)}$ to the amplitudes $r_A^{(l,r)}$ of the Andreev
backscattering process \cite{216,217}. Let us consider for example
the Andreev backscattering of an electron at the left interface
(Fig.~2).
\begin{figure}
\begin{center}
\includegraphics[angle=0, width=9cm]{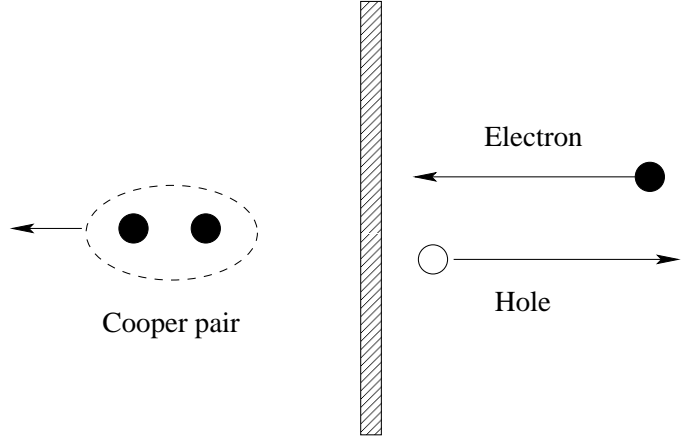}
\caption{A schematic picture of Andreev reflection.}
\end{center}
\end{figure}
This process can be described as the annihilation of two electrons
with opposite momenta and spin projections at $x=0$. The
corresponding Hamiltonian is $\;h_A\sim r_A^{\ast
(l)}a_{p,\uparrow}a_{-p,\downarrow}$, or equivalently in the
coordinate representation $h_A\sim r_A ^{\ast
(l)}\Psi_{R\uparrow}(0) \Psi_{L\downarrow}(0)$. Here $r_A$ is the
amplitude of Andreev backscattering at the left interface,
\begin{equation}\label{29}
r_A^{(l)}=\frac{|t^{(l)}|^2 e^{\imath(\varphi_l+
\pi/2)}}{\sqrt{|t^{(l)}|^4+ 4|r^{(l)}|^4}} \,,
\end{equation}
$t^{(l)}$ is the transmission amplitude
($|t^{(l)}|^2+|r^{(l)}|^2=1$) and $\varphi_l$ is the phase of
superconducting order parameter at the left bank. An analogous
expression holds for the right interface. Notice that for a tunnel
junction $|t^{(l,r)}|\ll 1$ the amplitude of Andreev
backscattering is small - it is proportional to the transparency
$D_{l,r}\equiv|t^{(l,r)}|^2\ll 1$ of the barrier at the right
(left) interface. So in our model the effective boundary pairing
potential is
\begin{equation}\label{210}
\Delta_B^{(l)}=C\hbar v_F r_A ^{\ast (l)}, \,
\Delta_B^{(r)}=-C\hbar v_F r_A ^{\ast (r)} \,,
\end{equation}
where $C$ is a numerical factor which will be specified later.

\subsection{2.2.1. Tunnel Junction}
For poorly transmitting interfaces $D_{r,l}\ll 1$ the amplitude of
Andreev backscattering is small and we can  use perturbation
theory when evaluating the phase dependent part of the ground
state energy. In second order perturbation theory the ground state
energy takes the form
\begin{equation}\label{211}
\delta E^{(2)}(\varphi)=\sum_j
\frac{|<j|H_B^{(A)}|0>|^2}{E_0-E_j}=
\frac{1}{\hbar}\int_0^\infty d\tau
<0|H_B^{(A)\dag}(\tau) H_B^{(A)}(0)|0> \,.
\end{equation}
Here $H_B^{(A)}(\tau)$ is the boundary Hamiltonian (\ref{27}) in
the imaginary time Heisenberg representation. After substituting
Eq.~(\ref{27}) into  Eq.~(\ref{211}) we get the following formula
for $\delta E^{(2)}$ expressed in terms of electron correlation
function
\begin{eqnarray}\label{212}
&&\hspace{.8in}\delta E^{(2)}(\varphi)= -4C \hbar v_F^2 \,
\text{Re} ( r_A ^{\ast (l)}  r_A ^{(r)} \times \nonumber \\
&&\int_0^\infty d\tau [ <\Psi_{R \uparrow} (\tau,0) \Psi_{L
\downarrow} (\tau,0) \Psi_{L \downarrow}^\dag (0,L) \Psi_{R
\uparrow}^\dag (0,L) > + <\uparrow \Longleftrightarrow \downarrow>
] ) \,.
\end{eqnarray}

We will calculate the electron correlation function by making use
of the bosonization technique. The standard bosonisation formula
reads
\begin{equation}\label{213}
\Psi_{\eta,\sigma}(x,t)=\frac{1}{\sqrt{2\pi a}}\exp
\{\imath\eta\sqrt{4\pi}\Phi_{\eta,\sigma}(x,t)\}\,,
\end{equation}
where $a$ is the cutoff parameter ($a\sim \lambda_F$),
$\eta=(R,L)\equiv (1,-1), \sigma=(\uparrow,\downarrow)
\equiv(1,-1)$. The chiral bosonic fields in Eq.~(\ref{213}) are
represented as follows (see e.g. Ref.~[\onlinecite{214}])
\begin{equation}\label{214}
\Phi_{\eta,\sigma}(x,t)= \frac{1}{2} \hat \varphi_{\eta,\sigma} +
\hat \Pi_\sigma \frac{x-\eta v t}{L} + \varphi_{\eta,\sigma}
(x,t)\,.
\end{equation}
Here the zero mode operators $\hat \varphi_{\eta,\sigma},\hat
\Pi_\sigma $ obey the standard commutation relations for
''coordinate" and ''momentum" $[\hat \varphi_{\eta,\sigma},\hat
\Pi_{\sigma'}]= -\imath \eta\delta_{\sigma,\sigma'}$. They are
introduced for a finite length LL to restore correct canonical
commutation relations for bosonic fields \cite{218, 219}. Notice
that the topological modes associated with these operators fully
determine the Josephson current in a transparent ($D=1$) SLLS
junction \cite{225}. The nontopological components
$\varphi_{\eta,\sigma} (x,t)$ of the chiral scalar fields are
represented by the series
\begin{equation}\label{214a}
\varphi_{\eta,\sigma} (x,t)=\sum_q \frac{1}{\sqrt{2qL}}\left[
e^{\imath q(\eta x - v t)}\hat b_q + h.c.\right]\,,
\end{equation}
where $\hat b_q (\hat b_q^\dag)$ are the standard bosonic
annihilation (creation) operator; $L$ is the length of the
junction, $v$ is the velocity.

It is convenient here to introduce \cite{214} the charge
($\varrho$) and spin ($\sigma$) bosonic fields $\varphi_\sigma,
\theta_\varrho$, which are related to above defined chiral fields
$\varphi_{\eta,\sigma} $ by simple linear equation
\begin{equation}\label{216}
\left(
\begin{array}{c}
  \varphi_\sigma \\
  \theta_\varrho \\
\end{array}
\right) =\frac{1}{\sqrt{2}}(\varphi_{R\uparrow}\pm
\varphi_{L\uparrow}\mp
\varphi_{R\downarrow}-\varphi_{L\downarrow})
\end{equation}
(the upper sign corresponds to $\varphi_\sigma$ and the lower sign
denotes $\theta_\varrho$). After straightforward transformations
Eq.~(\ref{212}) takes the form
\begin{equation}\label{217}
\delta E^{(2)}(\varphi)= 4 C \hbar v_F^2 D \cos \varphi
\int_0^\infty d \tau \left[ \Pi_+(\tau)+ \Pi_-(\tau)\right]\,,
\end{equation}
where $D=D_l D_r \ll 1$ is the junction transparency and
\begin{eqnarray}\label{218}
&&\Pi_\pm (\tau)=(2\pi a^2)^{-2} \exp\{ 2\pi[
\ll\varphi_\sigma(\tau,-L) \varphi_\sigma\gg+\ll\theta_\varrho
(\tau,-L) \theta_\varrho\gg \pm \\
&&\pm \ll\theta_\varrho (\tau,-L) \varphi_\sigma\gg \pm
\ll\varphi_\sigma (\tau,-L) \theta_\varrho\gg]\} Q_\pm
(\tau)\,.\nonumber
\end{eqnarray}
Here $\varphi_\sigma\equiv\varphi_\sigma (0,0),
\theta_\varrho\equiv \theta_\varrho (0,0)$ and the double brackets
$\ll ...\gg$ denote the subtraction of the corresponding vacuum
average at the points $(\tau, x)=(0,0)$. Notice that the
superconducting properties of a LL are determined by the
correlators of $\theta_\varrho$ and $\varphi_\sigma$ bosonic
fields unlike the normal conducting properties where the fields
$\theta_\sigma$ and $\varphi_\varrho$ play a dominant role. The
factors $Q_\pm (\tau)$ originate from the contribution of zero
modes,
\begin{equation} \label{219}
Q_\pm (\tau)=\exp \{ \frac{\pi}{2}<[\hat \Pi_\uparrow - \hat
\Pi_\downarrow \pm \frac{\imath v_F \tau}{L} (\hat \Pi_\uparrow +
\hat \Pi_\downarrow )]^2 > \}e^{\pi v_F \tau/L}\,.
\end{equation}
With the help of a Bogoliubov transformation the chiral bosonic
fields in Eq.~(\ref{216}) can be expressed in terms of
noninteracting plasmonic modes with known propagators (see e.g.
Ref.~[\onlinecite{214}]).
\begin{figure}
\begin{center}
\includegraphics[angle=0, width=12cm]{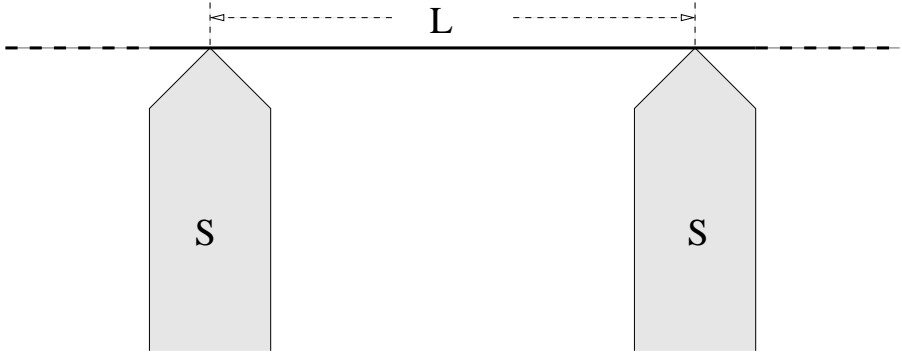}
\caption{A schematic picture of a SLLS junction formed by an
effectively infinite Luttinger liquid (LL) coupled to bulk
superconductors by side electrodes.}
\end{center}
\end{figure}
Two different geometries of SLLS junction have been considered in
the literature, viz., an effectively infinite LL connected by the
side electrodes to bulk superconductors \cite{220} (see Fig.~3)
and a finite LL wire coupled via tunnel barriers to
superconductors \cite{215,221}. Notice, that both model geometries
can be related to realistic contacts of a single wall carbon
nanotube with metals (see e.g. review \cite{222} and references
therein). The geometry of Fig.~3 could model the junction when
electron beam lithography is first used to define the leads and
then ropes of SWSN are deposited on top of the leads. A tunnel
junction of the type schematically shown in Fig.~4 is produced
when the contacts are applied over the nanotube rope.
\begin{figure}
\begin{center}
\includegraphics[angle=0, width=12cm]{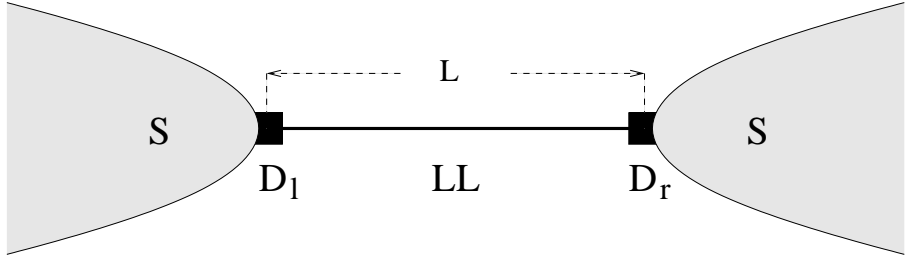}
\caption{A Luttinger liquid wire of length $L$ coupled to bulk
superconductors via tunnel barriers with transparencies
$D_{l(r)}$.}
\end{center}
\end{figure}

The topological excitations for an effectively infinite LL
($L\rightarrow\infty$) play no role and the corresponding
contributions can be omitted in Eqs.~(\ref{214}) and (\ref{218}),
$Q_\pm (\tau) \equiv 1.$ The propagators of noninteracting chiral
bosonic fields are (see e.g. \cite{214})
\begin{equation}\label{220}
\ll \varphi_{R/L,j}(t,x)\varphi_{R/L,k}\gg =-
\frac{\delta_{jk}}{4\pi} \ln\frac{a \mp x + s_k t}{a}\,,
\end{equation}
where $j,k =1,2$ and the plasmonic velocities $s_1= v_\varrho,
s_2= v_\sigma=v_F$ (see Eq.(\ref{26})). Finally the expression for
the Josephson current through a ''bulk-contacted" LL (Fig.~3)
takes the form \cite{220}
\begin{equation}\label{221}
J_{LL}^{(i)}= J_c ^{(0)} R_i(g_\varrho)\sin\varphi \,,
\end{equation}
where $J_c ^{(0)}= (Dev_F/L)(C/\pi)$ is the critical Josephson
current for noninteracting electrons, $R_i (g_\varrho)$ is the
interaction induced renormalization factor $(R_i (g_\varrho=1)=1)$
\begin{equation}\label{2211}
R_i (g_\varrho) = \frac{g_\varrho}{\sqrt{\pi}} \frac{\Gamma
(1/2g_\varrho)}{\Gamma (1/2+1/2g_\varrho)} F \left( \frac{1}{2},
\frac{1}{2}; \frac{1}{2g_\varrho}+\frac{1}{2}; 1-g_\varrho
^2\right) \left(\frac{a}{L}\right) ^{g_\varrho ^{-1}-1} \,.
\end{equation}
Here $g_\varrho$ is the correlation parameter of a spin-1/2 LL in
the charge sector Eq.~(\ref{26}), $\Gamma(x)$ is the gamma
function and $F(\alpha, \beta ; \gamma; z)$ is the hypergeometric
function (see e.g. Ref.~[\onlinecite{223}]). For the first time
the expression for $R_i (g_\varrho)$ in the integral form was
derived in Ref.~[\onlinecite{220}]. In the limit of strong
interaction $V_0/\hbar v_F\gg 1$ the renormalization factor is
small
\begin{equation}\label{222}
R_i (g_\varrho \ll 1) \simeq \frac{\pi}{2} \left(\frac{\hbar
v_F}{V_0}\right)^{3/2}\left(\frac{a}{L}\right)^{
\sqrt{\frac{2V_0}{\pi\hbar v_F}}} \ll 1\,,
\end{equation}
and the Josephson current through the SLLS junction is strongly
suppressed. This is nothing but a manifestation of the Kane-Fisher
effect \cite{22} in the Josephson current.

To evaluate the correlation function, Eq.~(\ref{218}), for a LL
wire of finite length coupled to bulk superconductors via tunnel
barriers, (Fig.~4), we at first have to formulate boundary
conditions for the electron wave function
\begin{equation}\label{223}
\Psi_\sigma (x) = e^{\imath k_F x}\Psi_{R,\sigma} (x)+ e^{-\imath
k_F x}\Psi_{L,\sigma} (x), \,\;\; \sigma=\uparrow,\downarrow
\end{equation}
at the interfaces $x=0,L$. To zeroth order of perturbation theory
in the barrier transparencies the electrons are confined to the
normal region. So the particle current $J_\sigma \sim \text {Re}
(\imath \Psi_\sigma^\ast\partial_{x}\Psi_\sigma )$ through the
interfaces is zero. For a single mode LL this requirement is
equivalent to the following boundary condition for the chiral
fermionic fields \cite{218,221}
\begin{equation}\label{224}
\Psi_{R,\sigma}^\ast (x) \Psi_{R,\sigma}(x)|_{x=0,L}=
\Psi_{L,\sigma}^\ast (x) \Psi_{L,\sigma}(x)|_{x=0,L}\,.
\end{equation}

These boundary conditions (LL with open ends) result in zero
eigenvalues of the ''momentum"-like zero mode operator $\hat
\Pi_\sigma$ and in the quantization of nontopological modes on a
ring with circumference $2L$ (see  Ref.~[\onlinecite{218}]). In
this case the plasmon propagators take the form
\begin{equation}\label{225}
\ll \varphi_{R/L,j}(t,x)\varphi_{R/L,k}\gg =-
\frac{\delta_{jk}}{4\pi} \ln\frac{1-e^{\imath \pi (\pm x- s_k
+\imath a)}}{\pi a/L}\,.
\end{equation}
With the help of Eqs.~(\ref{22}), (\ref{217})-(\ref{219}) and
Eq.~(\ref{225}) one readily gets the expression for the Josephson
current analogous to Eq.~(\ref{221}) $J_{LL}^{(f)}= J_c ^{(0)} R_f
(g_\varrho) \sin \varphi$, where now the critical Josephson
current of noninteracting electron is $J_c^{(0)}=
(Dev_F/4\pi)(C/\pi)$ and the renormalization factor
($R_f(g_\varrho=1)=1$) reads
\begin{equation}\label{2251}
R_f (g_\varrho) = \frac{2 g_\varrho^2}{2-g_\varrho^2} F \left(
\frac{2}{g_\varrho}; \frac{2}{g_\varrho}-g_\varrho;
\frac{2}{g_\varrho}-g_\varrho+1, -1\right) \left(\frac{\pi
a}{L}\right) ^{2(g_\varrho ^{-1}-1)} \,.
\end{equation}
Comparing $J_c^{(0)}$ with the well known formula for the critical
Josephson current in a low transparency SINIS junction (see e.g.
\cite{226}) we find the numerical constant $C=\pi$.

In the limit of strong interaction $g_\varrho\ll 1$
Eq.~(\ref{2251}) is reduced to the simple formula
\begin{equation}\label{226}
R_f (g_\varrho \ll 1)\simeq \frac{\pi}{2} \frac{\hbar
v_F}{V_0}\left(\frac{\pi a}{L}\right)^{2\sqrt{\frac{2V_0}{\pi\hbar
v_F}}}\ll 1\,.
\end{equation}
The dependence of the renormalization factor Eqs.~(\ref{2211}),
(\ref{2251}) on the strength of the electron-electron interaction
$V_0/\hbar v_F$ is shown in Fig.~5. The behavior of the Josephson
current as a function of the interaction strength is similar for
the two considered geometries. However we see that the interaction
influences the supercurrent more strongly for the case of
"end-coupled" LL wire.
\begin{figure}
\begin{center}
\includegraphics[angle=0, width=12cm]{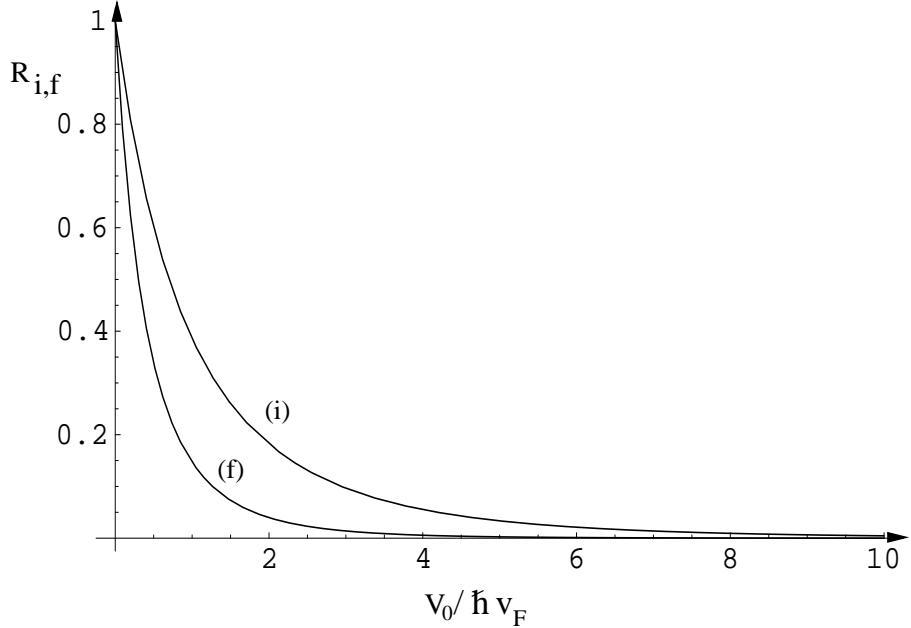}
\caption{Dependence of the renormalization factor $R_{i(f)}$ on
the dimensionless electron-electron interaction strength
$V_0/\hbar v_F$. Curve (i) corresponds to the case of
"side-coupled" LL wire, curve (f) to an "end-coupled" LL wire.}
\end{center}
\end{figure}

\subsection{2.2.2. Transparent Junction}

The case of perfectly transmitting interfaces in terms of the
boundary Hamiltonian (\ref{28}), (\ref{28}) formally corresponds
to the limit $V_B \rightarrow 0$ and not small $\Delta_B$. It can
not be perturbatively treated. Physically it means that charge is
freely transported through the junction and only pure Andreev
reflection takes place at the NS boundaries. It is well known that
at energies much smaller than the superconducting gap
($E\ll\Delta)$ the scattering amplitude of quasiparticles becomes
energy independent (see Eq.~(\ref{29})). This enable one to
represent the Andreev scattering process as a boundary condition
for a real space fermion operator. It was shown in Ref.~\cite{225}
that the corresponding boundary condition for chiral fermion
fields takes the form of a twisted periodic boundary condition
over the interval $2L$,
\begin{equation}\label{227}
\Psi_{L/R,\pm\sigma} (x\pm 2L, t)=e^{\pm\imath \vartheta}
\Psi_{L/R,\pm\sigma} (x,t)
\end{equation}
(the upper sign corresponds to the left-moving fermions, lower
sign - to right moving particles), where $\vartheta=\pi+\varphi$,
$\varphi$ is the superconducting phase difference and the phase
$\pi$ is acquired due to the Andreev reflection on two interfaces
(see e.g. Eq.~(\ref{29})). So the problem can be mapped \cite{225}
to the one for the persistent current of chiral fermions on a ring
of circumference $2L$. It is well known \cite{219, 2231} (see also
the review \cite{229}) that the persistent current in a perfect
ring (without impurities) in the continuum model does not depend
on the electron-electron interaction due to the translational
invariance of the problem. This "no-renormalization" theorem
allows us to conclude that the Josephson current in a perfectly
transmitting SLLS junction coincides with the supercurrent in a
one-dimensional long SNS ballistic junction \cite{26,230}
\begin{equation}\label{228}
J_{LL}=J_{\text{nonint}}=\frac{4eT}{\hbar}
\sum_{k=1}^{\infty}(-1)^{k+1}\frac{\sin k\varphi}{\sinh (2\pi
kT/\Delta_L)}\,,
\end{equation}
where $T$ is the temperature and $\Delta_L=\hbar v_F/L$. The
formal proof of this statement \cite{225} consist in evaluating
the partition function of the LL with the twisted boundary
conditions, Eq.~(\ref{227}), supplemented by a connection between
the $\Psi_{R,\sigma}$ and $\Psi_{L,\sigma}$ fields that follows
from the chiral symmetry. The superconducting phase difference
$\varphi$ couples only to zero modes of the charge current field
$\theta_\varrho$. In a Galilelian  invariant system zero modes are
not renormalized by the interaction and the partition function for
a SLLS junction exactly coincides with the one for a long SNS
junction.

We notice here that Eq.~(\ref{228}) holds not only for perfectly
transmitting interfaces. It also describes asymptotically at
$T\ll\Delta$ the Josephson current through a tunnel junction when
the interaction in the wire is assumed to be attractive. We have
seen already in the previous subsection that the electron-electron
interaction renormalizies the bare transparency of the junction
due to the Kane-Fisher effect. The renormalization is known to
suppress the electron current for a repulsive interaction and to
enhance it for an attractive forces \cite{22}. So one could expect
that for an attractive interaction the electron interface
scattering will be renormalized at low temperatures to perfect
Andreev scattering \cite{213}.

\section{2.3. Josephson current and the Casimir effect}

More then fifty years ago Casimir predicted \cite{231} the
existence of small quantum forces between grounded metallic plates
in vacuum. This force (a kind of Van der Waals force between
neutral objects) arises due to a change of the vacuum energy
(zero-point fluctuations) induced by the boundary conditions
imposed by the metallic plates on the fluctuating electromagnetic
fields (see e.g. Refs.~[\onlinecite{232,233}]). This force has
been measured (see e.g. one of the recent experiments \cite{234}
and the references therein) and in quantum field theory the
Casimir effect is considered as the most spectacular manifestation
of zero-point energy. In a general situation the shift of the
vacuum energy of fluctuating fields in a constrained volume is
usually called the Casimir energy $E_C$. For a field with zero
rest mass dimensional considerations result in a simple behavior
of the Casimir energy as a function of geometrical size. In 1D
$E_C\sim \hbar v/L$, where $v$ is the velocity. We will now show
that the Josephson current in a long SNS junction from a field
theoretical point of view can be considered as a manifestation of
the Casimir effect. Namely, the Andreev boundary condition changes
the energy of the "Fermi sea" of quasiparticles in the normal
region. This results in the appearance of: (i) an additional
cohesive force between the superconducting banks \cite{235}, and
(ii) a supercurrent induced by the superconducting phase
difference.

As a simple example we evaluate the Josephson current in a long
transparent 1D SNS junction by using a field theoretical approach.
Andreev scattering at the NS interfaces results in twisted
periodic boundary conditions, Eq.~(\ref{225}), for the chiral
fermion fields \cite{219}. So the problem is reduced to the
evaluation of the Casimir energy for chiral fermions on an $S^1$
manifold of circumference $2L$ with  "flux" $\vartheta$. Notice
that the left- and right-moving quasiparticles feel opposite (in
sign) "flux" (see Eq.~(\ref{227})). The energy spectrum takes the
form ($\Delta_L=\hbar v_F/L$)
\begin{equation}\label{229}
E_{n,\eta}(L,\varphi)= \pi\Delta_L
(n-\frac{1}{2}+\eta\frac{\varphi}{2\pi}),\, n=0,\pm 1, \pm 2, ...,
\eta=\pm 1\,,
\end{equation}
and coincides (as it should be) with the electron and hole
energies calculated by matching the quasiparticle wave functions
at the NS boundaries \cite{26}. The Casimir energy is defined as
the shift of the vacuum energy induced by the boundary conditions
\begin{equation}\label{230}
E_C(L,\varphi)=2 \left(-\frac{1}{2}\right) \left[\sum_{n,\eta}
E_{n,\eta}(L,\varphi) - \sum_{n,\eta}
E_{n,\eta}(L\rightarrow\infty)\right]\,.
\end{equation}
Notice, that the factor ($-1/2$) in Eq.~(\ref{230}) is due to the
zero-point energy of {\em chiral} fermions, the additional factor
of $2$ is due to spin degeneracy. Both sums in Eq.~(\ref{230})
diverge and one needs a certain regularization procedure to
manipulate them. One of the most efficient regularization methods
in the calculation of vacuum energies is the so called generalized
zeta-function regularization \cite{236}. For the simple energy
spectrum, Eq.~(\ref{229}), this procedure is reduced to the
analytical continuation of the infinite sum over $n$ in
Eq.~(\ref{230}) to the complex plane,
\begin{equation}\label{231}
E_C(\varphi)=-\pi \Delta_L \lim_{s\rightarrow -1}
\sum_{n=-\infty,\eta=\pm 1}^{\infty}(n+a_\eta)^{-s}=
-\pi\Delta_L\sum_{\eta=\pm 1}[\zeta(-1,a_{\eta})+
\zeta(-1,-a_{\eta})+a_{\eta}]\,,
\end{equation}
where $\zeta (s,a)$ is the generalized Riemann $\zeta$-function
\cite{223} and $a_\eta=(\pi+\eta \varphi)/2\pi$. Using an
expression for $\zeta (-n,a)$ in terms of Bernoulli polynomials
that is well-known from textbooks (see e.g.
Ref.~[\onlinecite{223}]) one gets the desired formula for the
Casimir energy of a 1D SNS junction as
\begin{equation}\label{232}
E_C= 2\pi\frac{\hbar v_F}{L}\left[
\left(\frac{\varphi}{2\pi}\right)^2-\frac{1}{12}\right],\,
|\varphi|\leq\pi\,.
\end{equation}
The Casimir force $F_C$ and the Josephson current $J$ at $T=0$ are
\begin{equation}\label{233}
F_C=-\frac{\partial E_C}{\partial \varphi}=\frac{E_C}{L}\,\,\,,\,\,\,\,
J=\frac{e}{\hbar} \frac{\partial E_C}{\partial\varphi}=\frac{e
v_F}{L} \frac{\varphi}{\pi},\,\,\,\,|\varphi|\leq\pi\,.
\end{equation}
the expression for the Josephson current coincides with the
zero-temperature limit of Eq.~(\ref{228}). The generalization of
the calculation method to finite temperatures is straightforward.
The additional cohesive force between two bulk metals induced by
superconductivity is discussed in Ref.~[\onlinecite{235}]. In this
paper it was shown that for a multichannel SNS junction this force
can be measured in modified AFM-STM experiments, where force
oscillations in nanowires were observed.

The calculation of the Casimir energy for a system of interacting
electrons is a much more sophisticated problem. In
Ref.~[\onlinecite{215}] this energy and the corresponding
Josephson current were analytically calculated for a special
exactly solvable case of double-boundary LL. Unfortunately the
considered case corresponds to the attractive regime of LLs
($g_\varrho =2$ in our notation, see Eq.~(\ref{26} )) and the
interesting results obtained in Ref.~[\onlinecite{215}] can not be
applied for electron transport in quantum wires fabricated in 2DEG
or in individual SWNTs were the electron-electron interaction is
known to be repulsive.

\section{3. The effects of Zeeman splitting and spin-orbit interaction
in SNS junctions}

In the previous section we considered the influence of
electron-electron interactions on the Josephson current  in a
S/QW/S junction. Although all calculations were performed for a
spin-1/2 Luttinger liquid model, it is readily seen that the spin
degrees of freedom in the absence of a magnetic field are
trivially involved in the quantum dynamics of our system. In
essence, they do not change the results obtained for spinless
particles. For noninteracting electrons spin only leads to an
additional statistical factor $2$ (spin degeneracy) in the
thermodynamic quantities. At the first glance spin effects could
manifest themselves in SLLS junctions since it is known that in LL
the phenomena of spin-charge separation takes place (see e.g.
Ref.~[\onlinecite{214}]). One could naively expect some
manifestations of this nontrivial spin dynamics in the Josephson
current. Spin effects for interacting electrons are indeed not
reduced to the appearance of statistical factor. However, as we
have seen already in the previous sections, the dependence of the
critical Josephson current on the interaction strength is
qualitatively the same for spin-1/2 and spinless Luttinger
liquids. So it is for ease of calculations a common practice to
investigate weak superconductivity in the model of spinless
Luttinger liquid (see e.g. Ref.~[\onlinecite{215}]).

Spin effects in the Josephson current become important in the
presence of a magnetic field, spin-orbit interactions or
spin-dependent scattering on impurities. At first we consider the
effects induced by a magnetic field. Generally speaking a magnetic
field influences both the normal part of the junction and the
superconducting banks. It is the last impact that determines the
critical Josephson current in short and wide junctions. The
corresponding problem was solved many years ago and one can find
the analytical results for a short and wide junction in a magnetic
field parallel to the NS interface e.g. in
Refs.~[\onlinecite{31,32}].

In this review we are interested in the superconducting properties
of junctions formed by a long ballistic quantum wire coupled to
bulk superconductors. We will assume that a magnetic field is
applied {\em locally} i.e. only to the normal part of the junction
(such an experiment could be realized for instance with the help
of a magnetic tip and a scanning tunneling microscope). In this
case the only influence of the magnetic field on the electron
dynamics in a single channel (or few channel) QW is due to the
Zeeman interaction. For noninteracting electrons the Zeeman
splitting lifts the double degeneracy of Andreev levels in an SNS
junction and results in a periodic dependence of the critical
Josephson current on magnetic field \cite{33}.

Interaction effects can easily be taken into account for a 1D SLLS
junction in a magnetic field by using bosonization techniques. The
term in the Hamiltonian $\hat{H}_Z$, which describes the
interaction of the magnetic field $\overrightarrow{B}$ with the
electron spin $\overrightarrow{S}(x)$ is in bosonized form  (see
e.g. Ref.~[\onlinecite{214}])
\begin{equation}\label{31}
\hat{H}_Z= - g_f \mu_B B_z\int dx S_z(x),\,
S_z(x)=\frac{1}{\sqrt{2\pi}}\partial_x \varphi_\sigma\,,
\end{equation}
where $g_f$ is the g-factor, $\mu_B$ is the Bohr magneton and the
scalar field $\varphi_\sigma$ is defined in Eq.~(\ref{216}). As is
easy to see, this interaction can be transformed away in the LL
Hamiltonian by a coordinate-dependent shift of the spin bosonic
field $\varphi_\sigma\Rightarrow\varphi_\sigma+ \Delta_z x /\hbar
v_F\sqrt{2\pi},\,\Delta_z=g_f\mu_B B$ is the Zeeman splitting. So
the Zeeman splitting introduces an extra $x$-dependent phase
factor in the chiral components of the fermion fields and thus the
Zeeman interaction can be readily taken into account (see e.g.
Ref.~[ \onlinecite{34}]) by a slight change of the bosonization
formula (\ref{213})
\begin{equation}\label{32}
\psi_{\eta,\sigma}^{(Z)}(x,t)=\exp \left(\imath K_{\eta,\sigma}
x\right) \psi_{\eta,\sigma} (x,t),\,\,\,\,
K_{\eta,\sigma}=\frac{\Delta_z}{4\hbar v_F} \eta\sigma,\,\,\,
\eta,\sigma=\pm 1.
\end{equation}
The phase factor appearing in Eq.~(\ref{32}) results in a periodic
dependence of the Josephson current on magnetic field. In the
presence of Zeeman splitting the critical current, say, for a
tunnel SLLS junction, Eq.~(\ref{221}), acquires an additional
harmonic factor $\cos (\Delta_Z/\Delta_L)$, the same as for
noninteracting particles.

\subsection{3.1. Giant Critical Current in a Magnetically Controlled
Tunnel Junction}

Interesting physics for low-transparency junctions appears when
resonant electron tunneling occurs. In this subsection we consider
the special situation when the conditions for resonant tunneling
through a junction are induced by superconductivity. The device we
have in mind is an SNINS ballistic junction formed in a 2DEG with
a tunable tunnel barrier ("I") and a tunable Zeeman splitting
which can be provided for instance with the help of a magnetic tip
and a scanning tunneling microscope (STM). In quantum wires
fabricated in 2DEG the effects of electron-electron interactions
are not pronounced and we will neglect them in what follows.

Resonant electron tunneling through a double barrier mesoscopic
structure is a well studied quantum phenomenon, which has numerous
applications in solid state physics. Recently a manifestation of
resonant tunneling in the persistent current both in
superconducting \cite{36} and in normal systems \cite{37} was
studied. In these papers a double-barrier system was formed by the
two tunnel barriers at the NS interfaces \cite{36} or in a normal
metal ring \cite{37}. It was shown that for resonance conditions
(realized for a special set of junction lengths \cite{36} or
interbarrier distances \cite{37}) a giant persistent current
appears which is of the same order of magnitude as the persistent
current in a system with only a single barrier. In the case of the
SINIS junction considered in Ref.~[\onlinecite{36}] the critical
supercurrent was found to be proportional to $\sqrt{D}$, where $D$
is the total junction transparency. Notice that the normal
transmission coefficient for a symmetric double-barrier structure
(i.e. the structure with normal leads) at resonance conditions
does not depend on the barrier transparency at all. It means that
for the hybrid structure considered in Ref.~[\onlinecite{36}] the
superconductivity actually suppresses electron transport.

Now we show \cite{38} that in a magnetically controlled single
barrier SFIFS junction ("F" denotes the region with nonzero Zeeman
splitting) there are conditions when superconductivity in the
leads strongly {\em enhances} electron transport. Namely, the
proposed hybrid SFIFS structure is characterized by a giant
critical current $J_c\sim\sqrt{D}$, ($D$ is the junction
transparency) while the normal conductance $G$ is proportional to
$D$.

For a single barrier SFIFS junction of length $L$, where the
barrier is located at a distance $l\ll L$ measured from the left
bank, the spectrum of Andreev levels is determined from the
transcendental equation \cite{38}
\begin{equation}\label{33}
\cos\frac{2E\pm\Delta_Z}{\Delta_L}+
R\cos\frac{2E\pm\Delta_Z}{\Delta_{L-2l}}+D\cos\varphi=0\,,
\end{equation}
where $\Delta_x=\hbar v_F/x$ and $D+R=1$, $\Delta_Z $ is the
Zeeman splitting. In the limit $\Delta_Z=0$ Eq.~(\ref{33}) is
reduced to a well-known spectral equation for Andreev levels in a
long ballistic SNS junction with a single barrier \cite{226,39}.

At first we consider the symmetric single-barrier junction, i.e.
the case when the scattering barrier is situated in the middle of
the normal region $l=L/2$. Then the second cosine term in the
spectral equation is equal to one and Eq.~(\ref{33}) is reduced to
a much simpler equation which is easily solved analytically. The
evaluation of the Josephson current shows \cite{38} that for $D\ll
1$ and for a discrete set of Zeeman splittings,
\begin{equation}\label{34}
\Delta_Z^{k}=\pi (2k+1)\Delta_L,\, k=0,1,2,...\,,
\end{equation}
the resonance Josephson current (of order $\sqrt{D}$) is
developed. At $T=0$ it takes the form
\begin{equation}\label{35}
J_r (\varphi)=\frac{e v_F}{L}\sqrt{D} \frac{\sin \varphi}{|\sin
(\varphi/2|)}\,.
\end{equation}
This expression has the typical form of a resonant Josephson
current associated with the contribution of a single Andreev level
(see Ref.~[ \onlinecite{226}]). One can interpret this result as
follows. Let us assume for a moment that the potential barrier in
a symmetric SNINS junction is infinite. Then the system breaks up
into two identical INS-hybrid structures. In each of the two
systems de Gennes-Saint-James energy levels with spacing
$2\pi\Delta_L$ are formed \cite{311}. For a finite barrier these
levels are split due to tunneling with characteristic splitting
energy $\delta\sim \sqrt{D}\Delta_L$. The split levels being
localized already on the whole length $L$ between the two
superconductors are nothing but the Andreev-Kulik energy levels
i.e. they depend on the superconducting phase difference. Although
the partial current of a single level is large ($\sim\sqrt{D}$)
(see Refs.~[\onlinecite{226,36}]), the current carried by a pair
of split levels is small ($\sim D$) due to a partial cancellation.
At $T=0$ all levels above the Fermi energy are empty and all
levels below $E_F$ are filled. So in a system without Zeeman
splitting the partial cancellation of pairs of tunnel-split energy
levels results in a small critical current ($\sim D$). The Zeeman
splitting $\Delta_Z$ of order $\Delta_L$ (see Eq.~(\ref{34}))
shifts two sets ("spin-up" and "spin-down") of Andreev levels so
that the Fermi energy lies in between the split levels. Now at
$T=0$ only the lower state is occupied and this results in an
uncompensated large ($\sim\sqrt{D}$) Josephson current. Since the
quantized electron-hole spectrum is formed by Andreev scattering
at the NS interfaces, the resonance structure for a single barrier
junction disappears when the leads are in the normal
(nonsuperconducting) state. So, the electron transport through the
normal region is enhanced by superconductivity. Electron spin
effects (Zeeman splitting) are crucial for the generation of a
giant Josophson current in a single barrier junction.

The described resonant transport can occur not only in  symmetric
junction. For a given value of Zeeman splitting $\Delta_Z^{(k)}$
from Eq.~(\ref{34}) there is a set of points \cite{38} (determined
by their coordinates $x_m^{(k)}$ counted from the middle of the
junction)
\begin{equation}\label{36}
x_m^{(k)}= \pm\frac{m}{2k+1}L
\end{equation}
($m$ is the integer in the interval $0\leq m\leq k+1/2$), where a
barrier still supports resonant transport. The temperature
dependence of the giant Josephson current is determined by the
energy scale $\delta\sim \sqrt{D}\Delta_L$ and therefore at
temperatures $T\sim \delta$, which are much lower then $\Delta_L$,
all resonance effects are washed out.

\subsection{3.2. Giant Magnetic Response of a Quantum Wire
Coupled to Superconductors}

It is known that the proximity effect produced in a wire by
superconducting electrodes strongly enhances the normal
conductance of the wire for certain value of the superconducting
phase difference (giant conductance oscillations \cite{312}). For
ballistic electron transport this effect has a simple physical
explanation \cite{313} in terms of Andreev levels. Consider a
multichannel ballistic wire perfectly (without normal electron
backscattering) coupled to bulk superconductors. The wire is
assumed to be connected to normal leads via tunnel contacts. In
the first approximation one can neglect the electron leakage
through the contacts and then the normal part of the considered
Andreev interferometer is described by a set of Andreev levels
produced by superconducting mirrors . When the distance $L$
between the mirrors is much longer then the superconducting
coherence length $L\gg \xi_0= \hbar v_F /\Delta$ ($\Delta$ is the
superconducting gap), the spectrum takes a simple form \cite{26}
\begin{equation}\label{37}
E_{n,\pm}^{(j)}=\frac{\hbar v_F^{(j)}}{2L}[\pi(2n+1)\pm
\varphi],\, n=0, \pm 1, \pm 2,...\,,
\end{equation}
where $v_F^{(j)}$ is the Fermi velocity of the j-th transverse
channel ($j=1, 2,...N_\bot$). It is evident from Eq.~(\ref{37})
that at special values of phase difference $\varphi_n=\pi(2n+1)$
energy levels belonging to different transverse channels $j$,
collapse to a single multi-degenerate ($N_\bot$) level exactly at
the Fermi energy. So resonant normal electron transport through a
multichannel wire (the situation which is possible for symmetric
barriers in the normal contacts) will be strongly enhanced at
$\varphi=\varphi_n$. The finite transparency of the barriers
results in a broadening and a shift of the Andreev levels. These
effects lead to a broadening of the resonance peaks in giant
conductance oscillations at low temperatures \cite{313}.

Magnetic properties of a quantum wire coupled to superconductors
can also demonstrate a behavior analogous to the giant conductance
oscillations. We consider a long perfectly transmitting SNS
junction in a local (applied only to the normal region) magnetic
field. In this case the only influence of the magnetic field on
the Andreev level structure is through the Zeeman coupling. The
thermodynamic potential $\Omega_A(\varphi, B)$ calculated for
Zeeman-split Andreev levels is \cite{235}
\begin{equation}\label{38}
\Omega_A(\varphi, B)= 4T \sum_{\{j\}}^{N_\bot}\sum_{k=1}^{\infty}
\frac{(-1)^k}{k}\frac{\cos k\varphi \cos k\chi_j}{\sinh (2\pi k
T/\Delta_L^{(j)})}\,.
\end{equation}
Here $\chi_j=\Delta_Z/\Delta_L^{(j)}, \Delta_Z =g\mu_B B$ is the
Zeeman energy splitting, $\Delta_L^{(j)}=\hbar v_F^{(j)}/L$ and
$v_F^{(j)}$ is the Fermi velocity in the $j$-th transverse
channel, $\{j\}$ is the set of transverse quantum numbers. In
Ref.~[\onlinecite{235}] the normal part of the SNS junction was
modelled by a cylinder of length $L$ and cross-section area
$S=V/L$. Hard-wall boundary conditions for the electron wave
function on the cylinder surface were assumed. Then the set
$\{j\}$ is determined by the quantum numbers ($l,n$) that label
the zeroes $\gamma_{l,n}$ of the Bessel function
$J_l(\gamma_{l,n})=0$ and the velocity $v_F^{(l,n)}$ takes the
form
\begin{equation}\label{39}
v_F^{(l,n)}=\sqrt{\frac{2}{m} \left( \varepsilon_F - \gamma_{ln}^2
\frac{\pi\hbar^2 L}{2mV}\right)}\,.
\end{equation}
It is evident from Eq.~(\ref{38}) that the
superconductivity-induced magnetization
\begin{equation}\label{310}
M_A=-\frac{\partial \Omega_{A}(\varphi,B)}{\partial B}
\end{equation}
at high temperatures ($T\gg \Delta_L$) is exponentially small and
does not contribute to the total magnetization of the junction. At
low temperatures $T\rightarrow 0$ the magnetization peaks at
$M_A\sim N_\bot g \mu_B$ where the superconducting phase
difference is an odd multiples of $\pi$ (see Fig.~6 which is
adapted from Ref.~[\onlinecite{235}]).
\begin{figure}
\begin{center}
\includegraphics[angle=0, width=12cm]{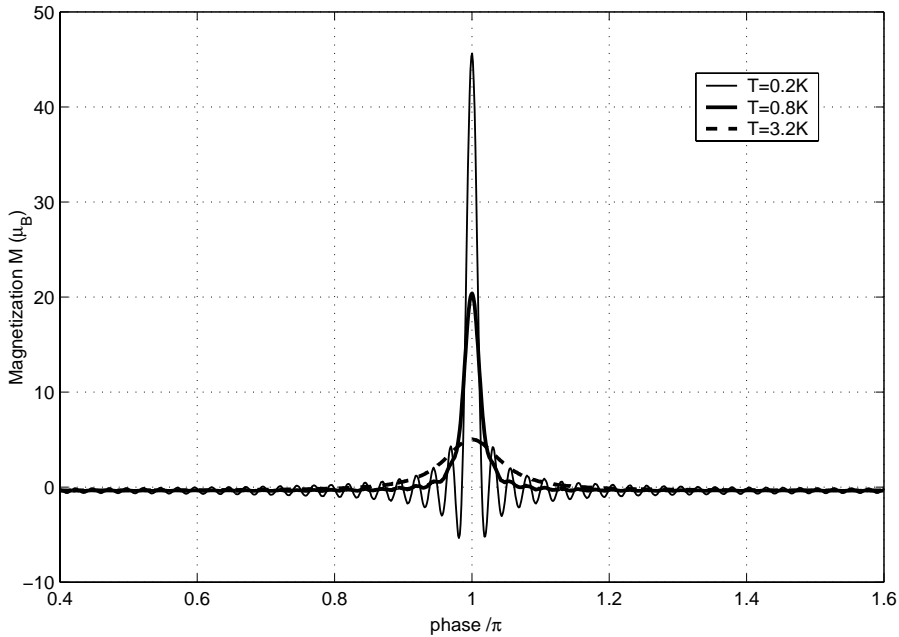}
\caption{Dependence of the magnetization $M$ of an SNS junction on
the superconducting phase difference for different temperatures.}
\end{center}
\end{figure}
The qualitative explanation of this resonance behavior of the
magnetization is as follows. It is known \cite{314} that for
$\varphi=\varphi_n\equiv (2n+1)\pi$ ($n$ is the integer) the two
Andreev levels $E_A^{(\pm)}=\pm\Delta_Z$ become $2N_\bot$-fold
degenerate. At $T\rightarrow 0$ the filled state $E_A^{(-)}$
dominates in the magnetization at $\varphi=\varphi_n$ since at
other values of superconducting phase the sets of Andreev levels
corresponding to different transverse channels contribute to
magnetization Eqs.~(\ref{38}), (\ref{310}) with different periods
in "magnetic phase" $\chi_j$ (i.e. in general, incoherently) and
their contributions partially cancel each other. Notice also that
for a fixed volume $V$, the number of transverse channels $N_\bot$
has a step-like dependence on the wire diameter. So at resonance
values of the phase difference $\varphi=\varphi_n$ one can expect
a step-like behavior of the magnetization as a function of wire
diameter \cite{235}. This effect is a magnetic analog of the
Josephson current quantization in a short SNS junction \cite{29}
considered in section 2.1.

\subsection{3.3. Rashba Effect and Chiral Electrons in Quantum Wires}

Another type of system where spin is nontrivially involved in the
quantum dynamics of electrons are conducting structures with
strong spin-orbit (s-o) interaction. It has been known for a long
time \cite{315} that the s-o interaction in the 2DEG formed in a
GaAs/AlGaAs inversion layer is strong due to the structural
inversion asymmetry of the heterostructure. The appearance in
quantum heterostructures of an s-o coupling linear in electron
momentum is now called the Rashba effect. The Rashba interaction
is described by the Hamiltonian
\begin{equation}\label{311}
H_{so}^{(R)} = \imath \alpha_{so}\left( \sigma_y
\frac{\partial}{\partial x}-\sigma_x \frac{\partial}{\partial
y}\right)\,,
\end{equation}
where $\sigma_{x(y)}$ are the Pauli matrices. The strength of the
spin-orbit interaction is determined by the coupling constant
$\alpha_{so}$, which ranges in a wide interval (1-10)$\times
10^{-10}$ eV$\times$cm for different systems (see e.g.
Ref.~[\onlinecite{316}] and references therein). Recently it was
experimentally shown \cite{317,318,319} that the strength of the
Rashba interaction can be controlled by a gate voltage
$\alpha_{so}(V_G)$. This observation makes the Rashba effect a
very attractive and useful tool in spintronics. The best known
proposal based on the Rashba effect is the spin-modulator device
of Datta and Das \cite{320}.

The spin-orbit interaction lifts the spin degeneracy of the 2DEG
energy bands at $\overrightarrow{p}\neq 0$ ($\overrightarrow{p}$
is the electron momentum). The Rashba interaction, Eq.~(\ref{311})
produces two separate branches for "spin-up" and "spin-down"
electron states
\begin{equation}\label{312}
\varepsilon
(\overrightarrow{p})=\frac{\overrightarrow{p}^2}{2m}\pm
\frac{\alpha_{so}}{\hbar} |\overrightarrow{p}|\,.
\end{equation}
Notice that under the conditions of the Rashba effect the electron
spin lies in a 2D plane and is always perpendicular to the
electron momentum. By the terms "spin-un"("spin-down") we imply
two opposite spin projections at a given momentum. The spectrum
(\ref{312}) does not violate left-right symmetry, that is the
electrons with opposite momenta ($\pm \overrightarrow{p}$) have
the same energy. Actually, the time reversal symmetry of the
spin-orbit interaction, Eq.~(\ref{311}), imposes less strict
limitations on the electron energy spectrum, namely,
$\varepsilon_\sigma (-\overrightarrow{p})=\varepsilon_{-\sigma}
(\overrightarrow{p})$ and thus, the Rashba interaction can in
principle break the chiral symmetry. In Ref. ~[\onlinecite{316}]
it was shown that in quasi-1D quantum wires formed in a 2DEG by a
laterally confining potential the electron spectrum is
characterized by a dispersion asymmetry $\varepsilon_\sigma
(-\overrightarrow{p})\neq \varepsilon_\sigma
(\overrightarrow{p})$. It means that the electron spectrum
linearized near the Fermi energy is characterized by two different
Fermi velocities $v_{1(2)F}$ and, what is more important,
electrons with large (Fermi) momenta behave as chiral particles in
the sense that in each subband  (characterized by Fermi velocity
$v_F^{(1)}$ or $v_F^{(2)}$) the direction of the electron motion
is correlated with the spin projection \cite{316, 322} (see
Fig.~7).
\begin{figure}
\begin{center}
\includegraphics[angle=0, width=12cm]{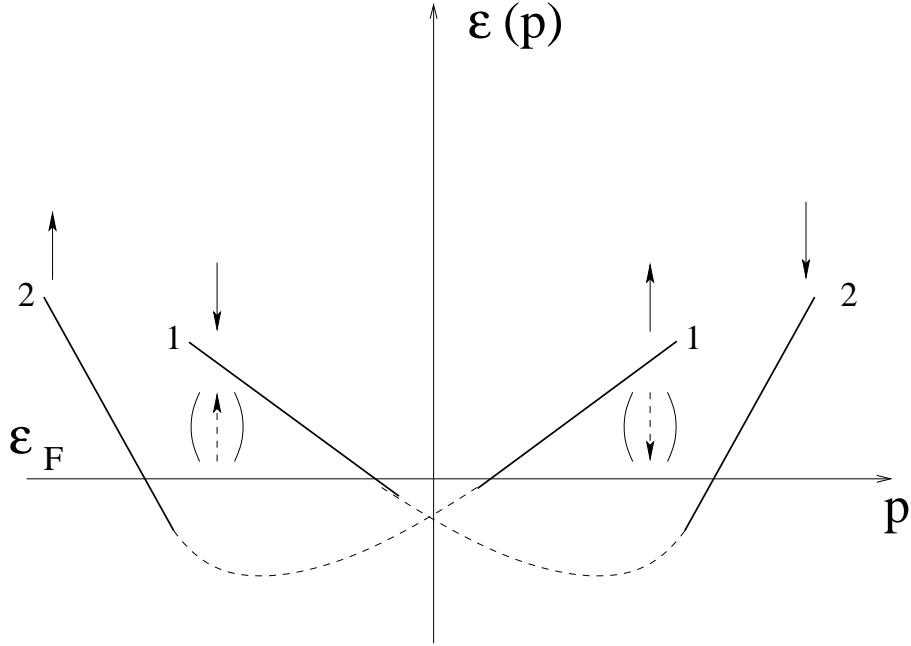}
\caption{Schematic energy spectrum of 1D electrons with dispersion
asymmetry. Particles with energies close to the Fermi energy
$\varepsilon_F$ have an almost linear dependence on momentum and
are classified by their Fermi velocities ($v_{1F}$-subband 1,
$v_{2F}$-subband 2). Solid line for spin projections correspond to
the case of weak s-o interaction; spin in parentheses indicates
the spin projections in subband 1 for strong Rashba interaction.}
\end{center}
\end{figure}
It is natural in this case to characterize the spectrum by the
asymmetry parameter
\begin{equation}\label{313}
\lambda_a=\frac{v_{1F}-v_{2F}}{v_{1F}+v_{2F}}\,,
\end{equation}
which depends on the strength of Rashba interaction
$\lambda_a(\alpha_{so}=0)=0$. The asymmetry parameter grows with
the increase of $\alpha_{so}$ and can be considered in this model
as the effective dimensional strength of the Rashba interaction in
a 1D quantum wire \cite{316}. Notice that the spectrum proposed in
Refs.~[\onlinecite{316,322}] (Fig.~7, solid lines for spin
projections) does not hold for strong s-o interactions, when
$\lambda_a$ is not small. Spin is not conserved in the presence of
the s-o interaction and the prevailing spin projection of electron
states in quasi 1D wires has to be independently calculated. It
was shown in Ref.~[\onlinecite{323}] by a direct calculation of
the average electron spin projection that for energies close to
$\varepsilon_F$ the electron spin projection for strong Rashba
interaction (comparable with the band splitting in the confining
potential) is strongly correlated with the direction of the
electron motion. Namely, the right-(R) and the left(L)-moving
electrons always have opposite spin projections regardless of
their velocities (see Fig.~7, where the parentheses indicate the
spin projection for strong Rashba interaction). For our choice of
Rashba s-o Hamiltonian, Eq.~(\ref{311}), "R"-electrons ($k_x>0$)
will be "down-polarized" ($<\sigma_y>=-1$) and "L"-electrons
($k_x<0$) will be "up-polarized" ($<\sigma_y>=+1$) to minimize the
main part of electron energy $\sim ( \hbar^2/2m)< k_x +\sigma_y m
\alpha_{so}/\hbar>^2$ in the presence of strong spin-orbit
interaction \cite{323}.

Chiral electrons in 1D quantum wire result in such interesting
predictions as "spin accumulation" in normal wires \cite{323} or
Zeeman splitting induced supercurrent in S/QW/S junction
\cite{38}.

\subsection{3.4. Zeeman Splitting Induced Supercurrent}

It was shown in the previous subsection that under the conditions
of the Rashba effect in 1D quantum wires the spin degree of
freedom is strongly correlated with the electron momentum. This
observation opens the possibility to magnetically control an
electric current. It is well known that in ring-shaped conductors
the current can be induced by magnetic flux due to the momentum
dependent interaction of the electromagnetic potential
$\overrightarrow{A}$ with a charged particle
$H_{\text{int}}=(e/mc) \overrightarrow{p} \overrightarrow{A}$.
Chiral properties of electrons in quasi-1D quantum wires allow one
to induce a persistent current via pure spin (momentum
independent) interaction $H=g \mu_B
\overrightarrow{S}\overrightarrow{H}$. Below we consider the
Josephson current in a ballistic S/QW/S junction in the presence
of Rashba spin-orbit interaction and Zeeman splitting. We will
assume at first that s-o interactions exist both in the normal
part of the junction and in the superconducting leads, so that one
can neglect the spin rotation accompanied by electron
backscattering induced by s-o interactions at the NS interfaces.
In other words the contacts are assumed to be fully adiabatic.
This model can be justified at least for a weak s-o interaction.
The energy spectrum of electrons in a quantum wire is shown in
Fig.~7 and the effect of the s-o interaction in this approach is
characterized by the dispersion asymmetry parameter $\lambda_a$,
Eq.~(\ref{313}).

For a perfectly transparent junction ($D=1$) the two subbands "1"
and "2" (see Fig.~7) contribute independently to the Andreev
spectrum which is described by two sets of levels \cite{38}
\begin{eqnarray}\label{314}
&&E_{n,\eta}^{(1)}=\pi\Delta_L^{(1)} \left( n+\frac{1}{2}+ \eta
\frac{\varphi+\chi_1}{2\pi}\right)\,, \\
&&E_{m,\eta}^{(2)}=\pi\Delta_L^{(2)} \left( m+\frac{1}{2}+ \eta
\frac{\varphi-\chi_2}{2\pi}\right)\,,\nonumber
\end{eqnarray}
where the integers $n,m=0, \pm 1, \pm 2, ...$ are ordinary quantum
numbers which label the equidistant Andreev levels in a long SNS
junction \cite{26}, $\eta=\pm 1,\, \Delta_L^{(j)}= \hbar v_{jF}/L
\,(j=1,2)$ and $\varphi$ is the superconducting phase difference.
The magnetic phases $\chi_j= \Delta_Z/\Delta_L^{(j)}$ characterize
the shift of Andreev energy levels induced by Zeeman interaction.
Notice that the relative sign between the superconducting phase
$\varphi$ and the magnetic phase $\chi_j$ is different for
channels "1" and "2". This is a direct consequence of the chiral
properties of the electrons in our model. In the absence of a
dispersion asymmetry ($v_{1F}=v_{2F}\equiv v_F$) the two sets of
levels in Eq.~(\ref{314}) describe the ordinary spectrum of
Andreev levels in a long transparent SFS junction ("F" stands for
the normal region with Zeeman splitting)
\begin{equation}\label{315}
E_{n,\eta,\sigma}=\pi\Delta_L \left( n+\frac{1}{2}+ \eta
\frac{\varphi}{2\pi}+ \sigma\frac{\chi}{2\pi}\right),\quad \quad
\eta,\sigma=\pm 1\,.
\end{equation}
Knowing explicitly the energy spectrum, Eq.~(\ref{314}), it is
straightforward to evaluate the Josephson current. It takes the
form \cite{38}
\begin{equation}\label{316}
J(\varphi,T,\Delta_Z)=\frac{2eT}{\hbar} \sum_{k=1}^\infty
(-1)^{k+1} \left[ \frac{\sin k(\varphi+\chi_1)}{\sinh (2\pi k
T/\Delta_L^{(1)})} + \frac{\sin k(\varphi-\chi_2)}{\sinh (2\pi k
T/\Delta_L^{(2)})}\right]\,.
\end{equation}
Here $T$ is the temperature. The formal structure of
Eq.~(\ref{316}) is obvious. The two sums in Eq.~(\ref{316})
correspond to the contributions of magnetically shifted sets of
levels "1" and "2" in Eq.~(\ref{314}). In the absence of any s-o
interaction the Zeeman splitting results only in an additional
$\cos (k\Delta_Z/\Delta_L)$ factor in the standard formula for the
supercurrent through a perfectly transmitting long SNS junction
\cite{230}. The most striking consequence of Eq.~(\ref{316}) is
the appearance of an \textit{anomalous} Josephson current
$J_{an}\equiv J(\varphi=0)$, when both the Zeeman splitting
($\Delta_Z$) and dispersion asymmetry ($\lambda_a$) are nonzero.
At high temperatures $T\geq \Delta_L^{(j)}$ the anomalous
supercurrent is exponentially small. In the low temperature regime
$T\ll \Delta_L^{(j)}$ it is a piece-wise constant function of the
Zeeman energy splitting $\Delta_Z$,
\begin{equation}\label{317}
J_{an}(\Delta_Z)=\frac{e}{\pi L}\sum_{k=1}^\infty
\frac{(-1)^{k+1}}{k} \left[ v_{1F}\sin \left( k
\frac{\Delta_Z}{\Delta_L^{(1)}}\right) - v_{2F}\sin \left( k
\frac{\Delta_Z}{\Delta_L^{(2)}}\right) \right]\,.
\end{equation}
For rational values $v_{1F}/v_{2F}=p/q$ ($p\leq q$ are the
integers) $J_{an}$ is a periodic function of the Zeeman energy
splitting with period $\delta \Delta_Z=2\pi q \Delta_L^{(1)}$,
otherwise it is a quasiperiodic function.
\begin{figure}
\begin{center}
\includegraphics[angle=0, width=12cm]{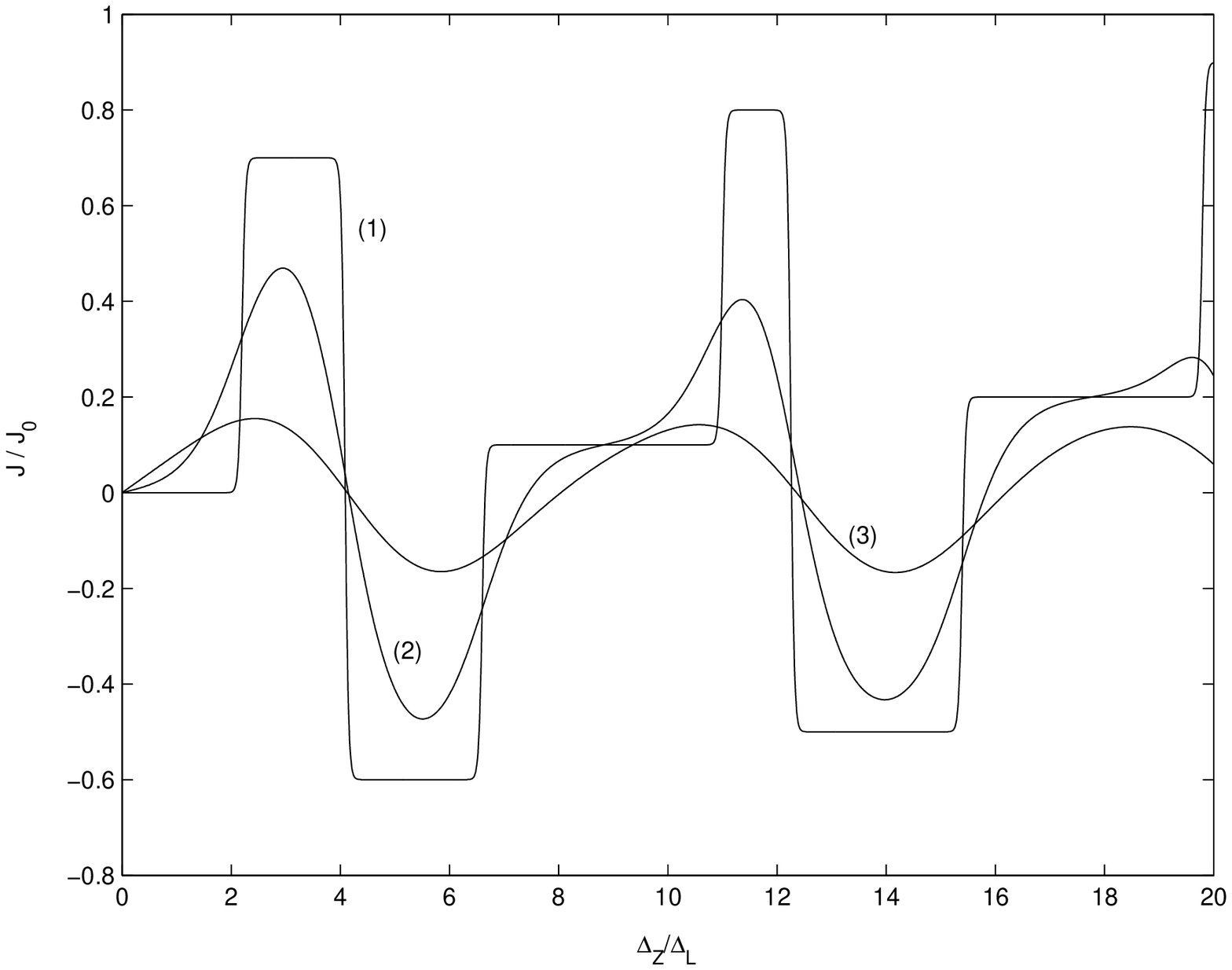}
\caption{Dependence of the normalized anomalous Josephson current
$J_{an}/J_0$} ($J_0=ev_F/L$) on the dimensionless Zeeman splitting
$\Delta_z/\Delta_L$ ($\Delta_L=\hbar v_F/L$) for asymmetry
parameter $\lambda_a=0.1$. The different plots (1-3) correspond to
different temperatures $T=(0.1, 1.5, 3.5)T^\ast$, where $T^\ast
=\Delta_L/2\pi$.
\end{center}
\end{figure}
The dependence of the normalized supercurrent $J_{an}/J_0\,$ (here
$J_0=ev_F/L,\, v_F=(v_{1F}+v_{2F})/2$) on the dimensionless Zeeman
splitting $\chi\equiv \Delta_Z/\Delta_L$ for $\lambda_a=0.1$ and
for different temperatures is shown in Fig.~8. We see that at
$T=0$ the Zeeman-splitting induced supercurrent appears abruptly
at finite values of $\Delta_Z$ of the order of the Andreev level
spacing.

Let us imagine now the situation when the Zeeman splitting arises
due to a local magnetic field (acting only on the normal part of
the junction) in the 2D plane applied normal to the quantum wire.
Then the vector product of this magnetic field and the electric
field (normal to the plane), which induces the Rashba interaction
determines the direction of the anomalous supercurrent. In other
words the change of the sign of the s-o interaction in
Eq.~(\ref{311}) or the sign of $\Delta_Z$ makes the supercurrent
Eq.~(\ref{317}) change sign as well.

Now we briefly discuss the case of a strong Rashba interaction
(the characteristic momentum $k_{so}=m/\hbar \alpha_{so}(V_g)$ is
of the order of the Fermi momentum). The electrons in a quantum
wire with strong Rashba coupling are chiral particles, that is the
right- and left-moving particles have opposite spin projections
\cite{323}. There is no reason to assume a strong s-o interaction
in 3D superconducting leads. We will follow the approach taken in
Refs.~[\onlinecite{325,323}], where the system was modelled by a
quantum wire  ($\alpha_{so}\neq 0$) attached to semi-infinite
leads with $\alpha_{so} =0$. In this model the SN interface acts
as a special strong scatterer where backscattering is accompanied
by spin-flip process. For a general nonresonant situation the
dispersion asymmetry is not important in the limit of strong
Rashba interaction and we can put $v_{1F}\approx v_{2F}\approx
v_F$. Then the Josephson current at $T=0$ up to numerical factor
takes the form
\begin{equation}\label{318}
J(\varphi, \Delta_Z)\approx D_{eff}(\alpha_{so}) \frac{ev_F}{L}
\sin \left( \varphi + \frac{\Delta_Z}{\Delta_L} \right)\,.
\end{equation}
Here $ D_{eff}(\alpha_{so})\ll 1$ is the effective transparency of
the junction. It can be calculated by solving the transition
problem for the corresponding normal junction \cite{323}. Anyway,
in the considered model for NS interfaces (nonadiabatic switching
on the Rashba interaction) even in the limit of strong Rashba
interaction the anomalous supercurrent $J_{an}=J(\varphi =0,
\Delta_Z)$ is small because of smallness of the effective
transparency of the junction. One could expect large current only
for special case of resonant transition. This problem has not yet
been solved.

\section{4. Conclusion}

The objective of our brief review was to discuss those
qualitatively new features of the Josephson effect that appear in
S/QW/S hybrid structures. Quantum wires are characterized by a 1D
or quasi-1D character of the electron conductivity. Electron
transport along QWs is ballistic and due to the weak screening of
the Coulomb interaction in 1D it is described by a Luttinger
liquid theory. So the first question we would like to answer was
--- what is the Josephson effect in SLLS junction? It was shown that
although electrons do not propagate in a LL weak link the
supercurrent in a perfectly transmitting SLLS junction exactly
coincides with the one in an SNS junction \cite{225}. This "no
renormalization" theorem is analogous to the result known for a LL
adiabatically coupled to nonsuperconducting leads \cite{23}. For a
tunnel SILLIS junction the dc Josephson current is described by
the famous Josephson current-phase relation, however now the
effective transparency $D_{\text{eff}}\ll 1$ defined as $J=J_0
D_{\text{eff}} \sin \varphi$ (where $J_0= e v_F /L$) strongly
depends on the aspect ratio of the LL wire $d/L$ ($d\sim
\lambda_F$ is the width of the nanowire), temperature and
electron-electron interaction strength. This result \cite{220} is
a manifestation of the Kane-Fisher effect \cite{22} in mesoscopic
superconductivity. It was also interesting for us (and we hope for
the readers as well) to find a close connection, rooted in the
Andreev boundary conditions, between the physics of a long SNS
junction and the Casimir effect (see section 2.3.).

Qualitatively new behavior of the proximity induced supercurrent
in nanowires is predicted for systems with strong spin-orbit
interactions. The Rashba effect in nanowires results in the
appearance of chiral electrons \cite{323, 316} for which the
direction of particle motion along the wire (right or left) is
strongly correlated with the electron spin projection. For chiral
electrons the supercurrent can be magnetically induced via Zeeman
splitting. The interplay of Zeeman, Rashba interactions and
proximity effects in quantum wires leads to effects that are
qualitatively different from those predicted for 2D junctions
\cite{326}.

It is worthwhile to mention here another important trend in
mesoscopic superconductivity, namely, the fabrication and
investigation of superconductivity-based qubits. Among different
suggestions and projects in this rapidly developing field, the
creation of a so-called single-Cooper-box (SCPB) was a remarkable
event \cite{327}. The SCPB consist of an ultrasmall
superconducting dot in tunneling contact with a bulk
superconductor. A gate electrode, by lifting the Coulomb blockade
of Cooper-pair tunneling, allows the delocalization of a single
Cooper pair between the two superconductors. For a nanoscale grain
the quantum fluctuations of the charge on the island are
suppressed due to the strong charging energy associated with a
small grain capacitance. By appropriately biasing the gate
electrode it is possible to make the two states on the dot,
differing by one Cooper pair, have the same energy. This two-fold
degeneracy of the ground state brings about the opportunity to
create a long-lived coherent mixture of two ground states (qubit).

The superconducting weak link which includes a SCPB as a tunnel
element could be very sensitive to external ac fields. This
problem was studied in \cite{328}, where the resonant microwave
properties of a voltage biased single-Cooper-pair transistor were
considered. It was shown that the quantum dynamics of the system
is strongly affected by interference between multiple
microwave-induced inter-level transitions. As a result the
magnitude and the direction of the dc Josephson current are
extremely sensitive to small variations of the bias voltage and to
changes in the frequency of the microwave field. This picture,
which differs qualitatively from the famous Shapiro effect
\cite{i8}, is a direct manifestation of the role the strong
Coulomb correlations play in the nonequilibrium superconducting
dynamics of mesoscopic weak links.

\subsection{Acknowledgment}
The authors thank E.~Bezuglyi, L.~Gorelik, A.~Kadigrobov and
V.~Shumeiko for numerous fruitful discussions. IVK and SIK
acknowledge the hospitality of the Department of Applied Physics
at Chalmers University of Technology and G\"{o}teborg University.
Financial support from the Royal Swedish Academy of sciences
(SIK), the Swedish Science Research Council (RIS) and the Swedish
Foundation for Strategic Research (RIS, MJ) is gratefully
acknowledged.

\end{document}